\documentclass[sigconf]{acmart}

\copyrightyear{2026}
\acmYear{2026}
\setcopyright{cc}
\setcctype{by-nc-nd}
\acmConference[WWW Companion '26]{Companion Proceedings of the ACM Web Conference 2026}{April 13--17, 2026}{Dubai, United Arab Emirates}
\acmBooktitle{Companion Proceedings of the ACM Web Conference 2026 (WWW Companion '26), April 13--17, 2026, Dubai, United Arab Emirates}
\acmPrice{}
\acmDOI{10.1145/3774905.3794661}
\acmISBN{979-8-4007-2308-7/2026/04}

\usepackage{amsmath}
\usepackage{amsthm}
\usepackage{amssymb}
\usepackage{mathtools}
\usepackage{xspace}
\usepackage[noend]{algorithmic}
\usepackage[ruled,vlined]{algorithm2e}
\usepackage{url}
\usepackage{makeidx}
\usepackage{enumerate}
\usepackage{epstopdf}
\usepackage{booktabs}
\usepackage{color}
\usepackage{thm-restate}
\usepackage{scalerel,stackengine}
\usepackage[shortlabels]{enumitem}
\usepackage{xr}
\usepackage{fancyvrb}
\usepackage{xcolor}
\usepackage[table]{xcolor}
\usepackage[most]{tcolorbox}
\usepackage{bold-extra}
\usepackage{subfigure}
\usepackage{float}
\usepackage{alltt}
\usepackage{soul}
\usepackage{fancyvrb}
\usepackage{multirow}
\usepackage{hyperref}
\usepackage[bottom]{footmisc}
\usepackage{bbding}
\usepackage{pifont}
\usepackage{arydshln}
\usepackage{tikz}
\usepackage{bm}
\usetikzlibrary{matrix, arrows}
\usepackage{balance}
\usepackage{pdfpages}

\definecolor{darkred}{HTML}{B32B3E}
\definecolor{darkblue}{HTML}{064A6C}

\newcommand\numberthis{\addtocounter{equation}{1}\tag{\theequation}}

\begin{document}

\title[RecoWorld for Agentic Recommender Systems]{RecoWorld: Building Simulated Environments for \\Agentic Recommender Systems}

\settopmatter{authorsperrow=4}
\author{Fei Liu}
\email{feiliu1@meta.com}
\affiliation{
  \institution{Meta Platforms, Inc.}
  \city{Menlo Park, CA}
  \country{USA}}

\author{Xinyu Lin}
\authornote{Work completed during internship at Meta.}
\email{xylin1028@gmail.com}
\affiliation{
  \institution{National University of Singapore}
  \city{}
  \country{}}

\author{Hanchao Yu}
\email{hanchaoyu@meta.com}
\affiliation{
  \institution{Meta Platforms, Inc.}
  \city{Menlo Park, CA}
  \country{USA}}

\author{Mingyuan Wu}
\authornotemark[1]
\email{mw34@cs.illinois.edu}
\affiliation{
  \institution{University of Illinois}
  \city{Champaign, IL}
  \country{USA}}

\author{Jianyu Wang}
\affiliation{
  \institution{Meta Platforms, Inc.}
  \city{Menlo Park, CA}
  \country{USA}}

\author{Qiang Zhang}
\affiliation{
  \institution{Meta Platforms, Inc.}
  \city{Menlo Park, CA}
  \country{USA}}

\author{Zhuokai Zhao}
\affiliation{
  \institution{Meta Platforms, Inc.}
  \city{New York, NY}
  \country{USA}}

\author{Yinglong Xia}
\affiliation{
  \institution{Meta Platforms, Inc.}
  \city{Menlo Park, CA}
  \country{USA}}

\author{Yao Zhang}
\affiliation{
  \institution{Meta Platforms, Inc.}
  \city{New York, NY}
  \country{USA}}

\author{Weiwei Li}
\affiliation{
  \institution{Meta Platforms, Inc.}
  \city{New York, NY}
  \country{USA}}

\author{Mingze Gao}
\affiliation{
  \institution{Meta Platforms, Inc.}
  \city{Menlo Park, CA}
  \country{USA}}

\author{Qifan Wang}
\affiliation{
  \institution{Meta Platforms, Inc.}
  \city{Menlo Park, CA}
  \country{USA}}

\author{Lizhu Zhang}
\email{lizhu@meta.com}
\affiliation{
  \institution{Meta Platforms, Inc.}
  \city{Menlo Park, CA}
  \country{USA}}

\author{Benyu Zhang}
\email{byzhang@meta.com}
\affiliation{
  \institution{Meta Platforms, Inc.}
  \city{Menlo Park, CA}
  \country{USA}}

\author{Xiangjun Fan}
\authornote{Corresponding author.}
\email{maxfan@meta.com}
\affiliation{
  \institution{Meta Platforms, Inc.}
  \city{Menlo Park, CA}
  \country{USA}}

\renewcommand{\shortauthors}{Fei Liu et al.}

\begin{abstract}
We present \textsc{RecoWorld}, a blueprint for building simulated environments tailored to agentic recommender systems. Such environments give agents a proper training space where they can learn from errors without impacting real users. \textsc{RecoWorld} distinguishes itself with \textbf{a dual-view architecture: a simulated user and an agentic recommender engage in multi-turn interactions aimed at maximizing user retention.} The user simulator reviews recommended items, updates its mindset, and when sensing potential user disengagement, generates reflective instructions. The agentic recommender adapts its recommendations by incorporating these user instructions and reasoning traces, creating a dynamic feedback loop that actively engages users. This process leverages the exceptional reasoning capabilities of modern LLMs. We explore diverse content representations within the simulator, including text-based, multimodal, and semantic ID modeling, and discuss how multi-turn RL enables the recommender to refine its strategies through iterative interactions. \textsc{RecoWorld} also supports multi-agent simulations, allowing creators to simulate the responses of targeted user populations. It marks an important first step toward recommender systems where users and agents collaboratively shape personalized information streams. We envision new interaction paradigms where ``user instructs, recommender responds,'' jointly optimizing user retention and engagement.
\end{abstract}



\begin{teaserfigure}
  \label{fig:teaser}
  \vspace{-0.1in}
\end{teaserfigure}


\maketitle

\section{Introduction}
\label{sec:intro}

\begin{table}
\centering
\begin{small}
\begin{tabular}{l|p{3in}}
\hline
\multicolumn{2}{|c|}{\cellcolor{blue!15} \textbf{Example Use Cases of} \textsc{RecoWorld}}\\
\hline
\multicolumn{2}{c}{\scriptsize}\\

& \textbf{Evaluating RecSys' Instruction Following Ability}.\; Instruction-following capabilities are essential for agentic RecSys, much like they are for LLMs. E.g., a user might tell the recommender system their preferences through natural language or voice, e.g., ``\emph{There are too many recommendations about hairstyling; I wanna see something different but related,}'' or ``\emph{I'd like to watch more content that people in the San Francisco Bay Area are watching right now.}'' If the agentic RecSys adjusts its recommendations based on these instructions, we need a simulated environment to evaluate how well it performed. One approach is to measure the system's success in instruction-following by using simulated user feedback as a proxy. This approach is important because agentic recommenders will need to handle a wide range of user instructions. Our \textsc{RecoWorld} provides a promising way to estimate system performance without relying on ground-truth annotated data.\\
\multicolumn{2}{c}{\scriptsize}\\[-0.2em]

& \textbf{Enabling Creators to Experiment with Publishing Strategies}.\;\; A simulated environment helps creators explore different content strategies and reach their goals, e.g., growing followers, boosting views, or getting more likes. It lets them try bold, different approaches without risking real user experience. For example, they can test new topics and styles to attract fresh audiences, find the right posting frequency to keep followers interested without annoying them, or see how controversial opinions might be received. By mimicking user feedback, this environment lets creators test the waters safely. Simulated user profiles can be based on their current audience, a general population sample, or even specific target groups.\\
\multicolumn{2}{c}{\scriptsize}\\[-0.2em]

& \textbf{Supporting New and Marginal Users in Exploring Interests}.\;\; Algorithms such as the multi-armed bandit are frequently used in RecSys to find the right balance between exploring new user interests and exploiting known preferences. I.e., an agent chooses from a fixed set of actions (interests) to maximize cumulative reward over time. A challenge here is to \emph{explore without compromising user experience}. \textsc{RecoWorld} addresses this by collecting simulated user engagement metrics, such as watch time and clicks, which serve as ``pseudo-rewards'' to evaluate the relevance of each interest. Our simulator can seamlessly integrate with contextual bandit by simulating groups of users with similar features, such as age, location, and past engagements, to provide collective exploration feedback.\\
\multicolumn{2}{c}{\scriptsize}\\[-0.2em]

& \textbf{Building a Leaderboard for Agentic RecSys}.\;\; Many enterprise recommender systems are developed in-house, making it difficult to replicate their results. The challenge is even greater with Agentic RecSys, which can take on various forms. They share an ``agentic'' nature, e.g., an agent (or multiple agents) interacts with an environment; the agent takes actions and receives observations and rewards in return. However, each task may have a different setup, e.g., variations in environmental states and reward signals. \textsc{RecoWorld} provides a shared platform with a community leaderboard, allowing different systems to be compared fairly. We invite practitioners to contribute their agentic tasks to \textsc{RecoWorld} using a common API. They can receive feedback signals from our simulated users and make their tasks available for the community to participate in. \textsc{RecoWorld} thus accelerates RL development, organizes various recommendation tasks from simple to complex for curriculum learning, and even fosters an agent-agent collaborative ecosystem.\\
\multicolumn{2}{c}{\scriptsize}\\[-0.2em]

\end{tabular}
\end{small}
\caption{We provide four example use cases of \textsc{RecoWorld}, while recognizing that building simulated environments for agentic RecSys can go well beyond these applications.}
\label{tab:use-cases}
\end{table}

Recommender systems have long relied on offline metrics such as Recall@N, NDCG, and counterfactual policy evaluation, paired with online A/B tests~\citep{zhai2024actionsspeaklouderwords,deng2025onerecunifyingretrieverank}. Offline evaluation is based on historical user behavior, which can introduce \emph{exposure bias}. This means the system often reinforces known patterns instead of discovering new user interests. In contrast, online A/B tests provide valuable insights, yet they involve a slow feedback loop and must be handled carefully with real users. As agentic recommender systems emerge, there is an increasing demand for simulated online environments that use LLMs to replicate user feedback. These simulations allow recommender systems to aggressively test bold, radically different strategies without compromising user experience.

Simulated environments have accelerated the progress of reinforcement learning (RL) research. For instance, OpenAI Gym~\citep{brockman2016openaigym} provides a collection of benchmarks for comparing the performance of RL algorithms. More recently, \citet{park2023generativeagentsinteractivesimulacra} pioneered the use of LLMs to simulate human behavior by developing a small-town simulation of generative agents. AgentCompany~\citep{xu2025theagentcompanybenchmarkingllmagents} investigated a self-contained environment that mimics a software company. AgentSociety~\citep{piao2025agentsocietylargescalesimulationllmdriven} is a simulator with over 10k agents designed to study societal impacts, such as universal basic income. CRMArena-Pro~\citep{huang2025crmarenaproholisticassessmentllm} utilizes simulated LLM agents to evaluate diverse business scenarios. These studies have shown promising initial results in using LLMs to simulate human behavior, paving the way for creating simulated environments to evaluate agentic recommender systems.

An ideal simulator would replicate real-world user interactions in a controlled environment~\citep{deffayet2024sardinesimulatorautomatedrecommendation,corecco2024suberrlenvironmentsimulated}. It allows developers to refine recommendation algorithms before deploying them. The simulator may include a variety of user profiles drawn from different demographics to reflect the diversity of the actual user base. It would simulate realistic user behaviors, such as clicking, sharing, buying, and giving feedback, using historical data or even starting from scratch. It might also factor in contextual information such as time of day, location, and device type, which can influence user interactions. While feedback from simulated users may not perfectly replicate every interaction, it remains valuable by providing reward signals that reflect population-level preferences.

\begin{figure*}
\centering
\includegraphics[width=0.93\textwidth]{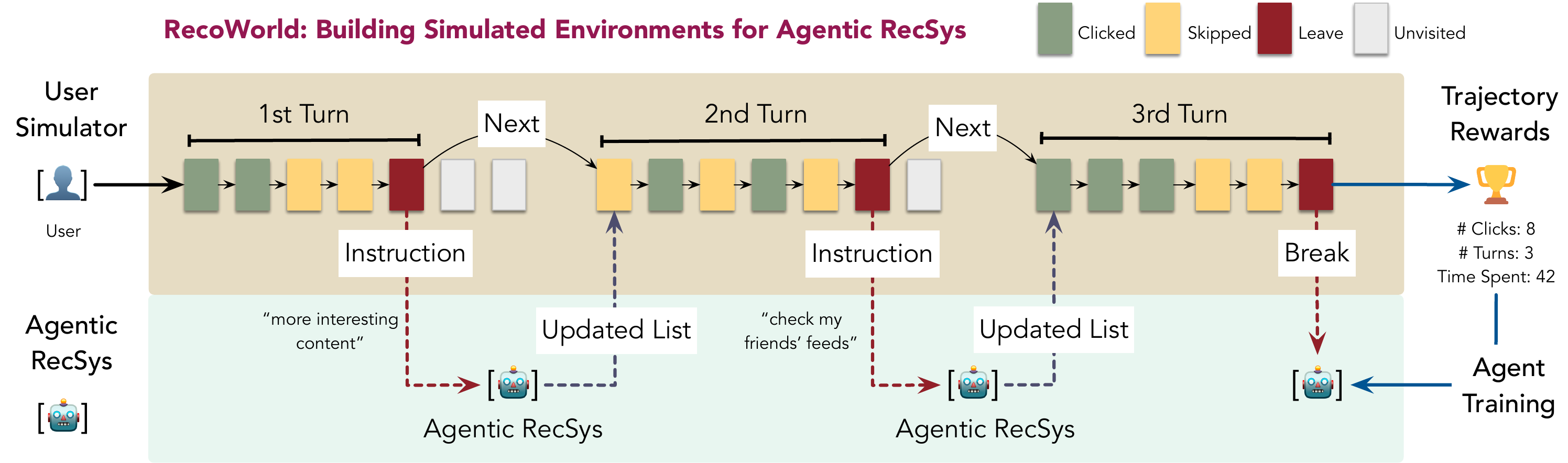}
\vspace{-0.1in}
\caption{A simulated user interacts with an agentic RecSys over multiple turns within a session.}
\label{fig:framework}
\vspace{-0.1in}
\end{figure*}

We introduce \textsc{RecoWorld}, a blueprint for building simulated environments that propel the development of agentic recommender systems~\citep{huang2025agenticrecommendersystemsera}. As these systems gain capabilities such as perception, reasoning, planning, memory, tool use, and autonomy, having environments for rapid system testing and refinement becomes essential before deploying them widely. We do not propose new agentic recommenders here; instead, our focus is on creating simulated environments that replicate user feedback. \textsc{RecoWorld} harnesses LLMs' strong reasoning, long-context understanding, and multimodal abilities to simulate user responses. It focuses on optimizing for long-term value through simulated interactions. The approach also emphasizes collective impacts, e.g., a post is distributed to a community of interconnected users rather than isolated individuals. The design of user simulation and reward signals in \textsc{RecoWorld} offers promising potential for enhancing existing recommender systems in modeling both short- and long-term user interests, as well as for generative recommenders~\citep{xia2025hierarchicaltreesearchbaseduser,wang2025gflowgrfinetuninggenerativerecommendation}. 

\vspace{0.05in}
\textbf{Our contributions in this paper are summarized as follows.}
\begin{itemize}[topsep=3pt,itemsep=0pt,leftmargin=*]

\item We introduce \textsc{RecoWorld}, a simulated environment where virtual users interact with agentic recommender systems to maximize engagement. \textsc{RecoWorld} generates multi-turn interaction trajectories, from which we extract engagement statistics that serve as reward signals for RL-based agent training.

\item Our user simulator models sequential user behavior. When a user is about to disengage, they can issue instructions, such as ``\emph{show me more interesting content}'', instead of simply dropping out. The agentic RecSys responds to these instructions by creating new item lists, aiming to re-engage the user. This cycle continues until the user chooses to exit without further instructions.

\item A successful simulator triggers accurate instructions that guide the recommender, while an effective recommender responds with item lists that enhance user satisfaction. To our knowledge, \textsc{RecoWorld} is the first environment to enable instruction-following recommenders in a Gym-like RL framework, marking a significant step toward more interactive recommender systems.

\end{itemize}

\vspace{-0.1in}
\section{Agentic Environments for RecSys}
\label{sec:agentic_env}

Agentic is emerging as a defining characteristic of next-generation foundation models. An agentic recommender system acts as an autonomous agent that actively learns from user interactions. It follows user instructions, acquires new skills to make proactive recommendations, and adapts its behavior based on these experiences~\citep{silver2025era}. This marks a shift from traditional recommender systems that primarily make passive suggestions~\citep{huang2025agenticrecommendersystemsera,shang2025agentrecbenchbenchmarkingllmagentbased,maragheh2025futureagenticdefinitionsperspectives}. With new capabilities to reason and interact with users, there comes the urgent need to develop new environments for evaluation. {Simulated environments} enable rapid system iteration, enhance data curation, and are increasingly leveraged for agent training. E.g., Claude uses simulated environments to check how the model responds to risky prompts, such as being asked to act as a dark web shopping assistant~\citep{anthropic2025claude}.

\textbf{Fundamental Components.}\quad Our \textsc{RecoWorld} utilizes a dual-view architecture. A \emph{user simulator} (a) generates simulated feedback by replicating the behavior of individual users or groups in response to recommendations, and (b) initiates requests to the recommender via explicit instructions or implicit signals. A second component is the \emph{recommender system} itself, which consists of modules for candidate retrieval, ranking, re-ranking, etc.; these modules work together to deliver personalized suggestions. An \textbf{agentic recommender system} extends this setup by involving one or multiple agents that interact with the simulated user. These agents may engage in conversation to clarify requests or better understand user preferences. When such a system can respond to user instructions and dynamically update recommendation lists to keep the user immersed in the session, it is referred to as an \textbf{instruction-following recommender}. Our focus is on developing robust user simulators, and we provide a concise overview of the Agentic RecSys architecture in \S\ref{sec:info_rec}.

\textbf{Multi-Turn Interactions in a Session.}\quad
A simulated user interacts with an agentic RecSys over multiple turns (see Figure \ref{fig:framework}). The user receives a list of recommended items and can take actions such as clicking or skipping. In the first turn, the user clicks on two items (green), skips two items (yellow), and then chooses to leave the session after the next item (red). Before leaving, the user reflects on their experience, identifies reasons for dissatisfaction, and gives a short instruction to the RecSys as feedback. If the RecSys cannot process user instructions, it simply updates recommendations based on recent interactions. If it can, it refreshes the recommendations to better address the user's feedback. The user reviews the new list, continues interacting, and may repeat the process of providing feedback if considering leaving. This cycle continues until the user decides to end the session without further feedback. The sequence of recommended items and the user's actions form the \textbf{interaction trajectory} for the session.

Consider an example instruction: User: ``\emph{Show me more interesting content.}'' System: ``\emph{Absolutely! Here are videos tailored to your interests and engagement history. Would you like to specify what you find interesting, such as trending topics, educational content, or entertainment? You can also tell me if you want more from specific creators.}'' \textcolor{black}{The agent perceives the user's explicit instruction and reasons over their engagement history, demographics, and behavior to create a configuration for the recommender system to execute. The system returns an updated recommendation list along with strategies to guide the user. The user simulator then generates feedback signals indicating whether the Agentic RecSys met the user's instructions and may provide the next instruction.}

\textbf{A successful simulator triggers accurate instructions that guide the recommender, while an effective recommender responds with item lists that enhance user satisfaction.} The user simulator and agentic RecSys collaboratively enhance session retention. We use trajectory-level interaction metrics such as total time spent as {reward signals} to measure the effectiveness of agentic RecSys. Following~\citet{brown2024largelanguagemonkeysscaling}, we generate \emph{multiple interaction trajectories} per user and initial recommendations. An LLM-based judge evaluates each trajectory against predefined task rubrics. Only trajectories that satisfy the success criteria are retained for training. Such creation of high-quality agentic trajectories, combined with general RL techniques that leverage reward signals and self-critique, is crucial for developing advanced agentic RecSys.

These reward signals are orthogonal to offline metrics. Metrics such as NDCG favor exploitation by optimizing for immediate relevance, while our simulator encourages exploration by modeling session dropout risk and optimizing for long-term user retention. We hypothesize that (a) high NDCG combined with high retention signals strong exploitation, with recommendations that are highly relevant; (b) high NDCG with low user retention reveals suboptimal recommendations; the content may be repetitive or insufficiently diverse, leading to user disengagement; and (c) low NDCG with high user retention likely suggests effective exploration, where users are exposed to novel content, supporting long-term retention.

\vspace{0.05in}
\noindent\textbf{Supporting Real and Simulated User Interactions.}\quad\;\;
(a) When simulated users are replaced by human annotators, the agentic RecSys interacts with actual users and receives authentic feedback signals, effectively functioning as an online experiment. (b) When Agentic RecSys is replaced with a traditional RecSys, the system ignores user instructions and generates recommendations only from the user's past engagement history, without adapting to additional feedback. (c) When instructions are disabled, the user simulator defaults to an evaluator that scores a list of items, without giving instructions to the RecSys for improvement. Since no improvement is requested, there will be no follow-up iterations; the session will end after a single turn.

\begin{figure*}
\centering
\includegraphics[width=6.2in]{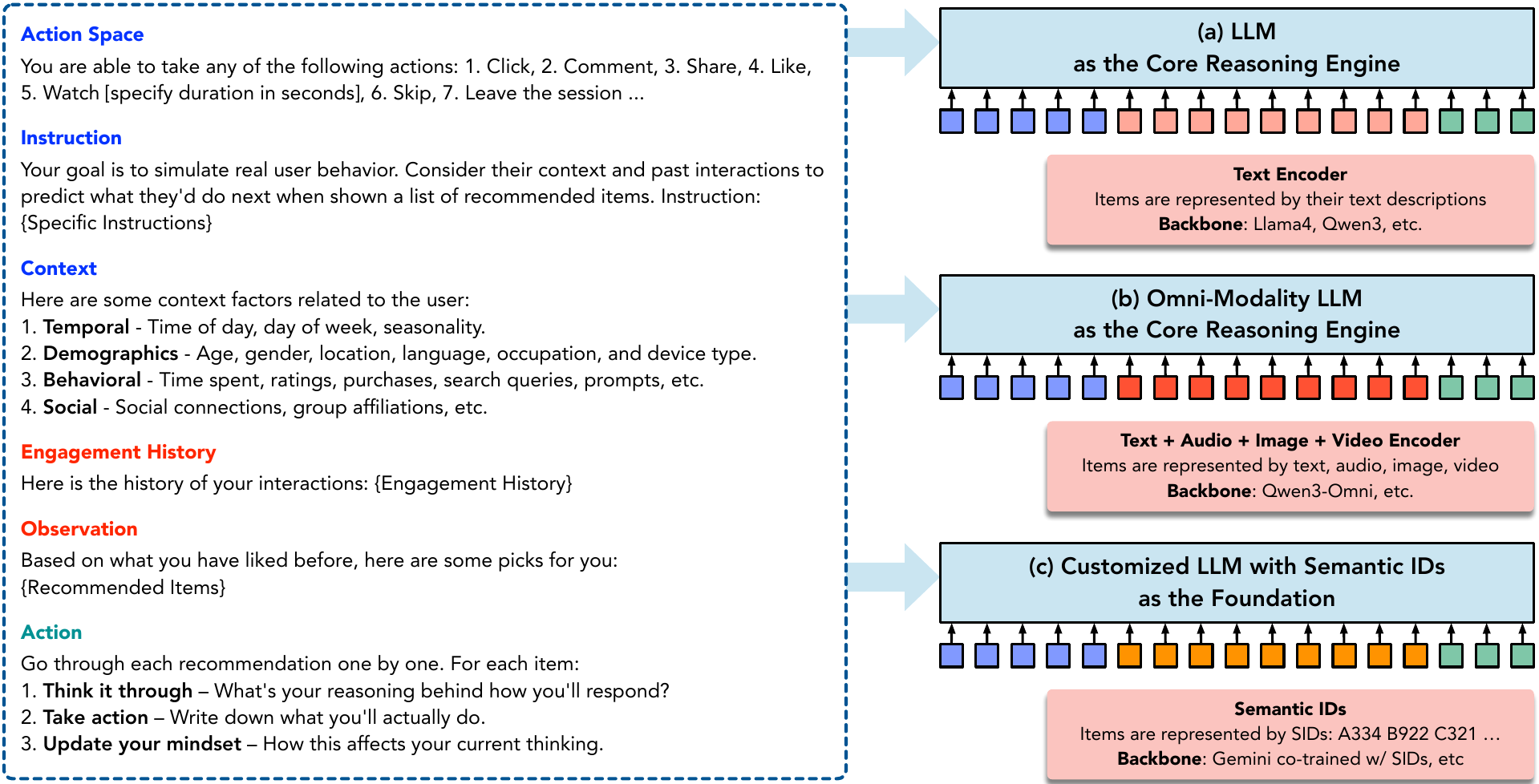}
\vspace{-0.05in}
\caption{Three modeling alternatives for engagement history that leverage LLMs' powerful reasoning capabilities (\S\ref{sec:user-sim}).}
\label{fig:example}
\vspace{-0.1in}
\end{figure*}

\textbf{Explicit Versus Implicit Instructions}\quad Importantly, user preferences are not conveyed solely through explicit instructions; they are also conveyed implicitly through behavior, whether users realize it or not. E.g., if a user frequently finishes watching videos that are >5 minutes, it indicates a receptiveness to long-form content, unlike those who prefer quick 10- or 20-second clips. If a user continues a session without interruption, it likely means they are satisfied with the recommendations. Conversely, if they request ``something more interesting,'' or if a user does not return for a while, it likely indicates that the previous recommendations failed to engage them. Implicit instructions can be inferred by monitoring user state. Users with similar behavioral patterns often share underlying implicit signals. Collaborative filtering can exploit these shared patterns to identify implicit instructions that reflect the collective interests of these user groups. Both implicit and explicit signals are crucial for agentic recommenders.

\textbf{Evaluating User Simulators.}\quad
Our paper does not present experimental results; instead, we outline evaluation designs to ensure user simulators are assessed in realistic settings. Our approach leverages existing RecSys datasets and validation by human annotators. When an annotator is available, they receive the same initial item list as the simulator and perform a single action on each item. The annotator can instruct the RecSys to update recommendations or end the session at any point. The system continues to provide updated recommendations in a multi-turn setting until the annotator requests a break without further instructions. We compare session-level interaction statistics between the simulated user and the human annotator as a sanity check for simulator effectiveness. Our user simulator can also be evaluated using existing recommendation datasets. The simulated user interacts with a recommended list (e.g., \{$i_1, i_2, i_3, i_4, i_5$\}), identifies break points (e.g., at $i_3$), and issues instructions to reorder the remaining items (\{$i_4, i_5$\} $\rightarrow$ \{$i_5, i_4$\}) to better match user instructions. This process continues until no further instructions are given, resulting in an updated list (e.g., \{$i_1, i_2, i_3, i_5, i_4$\}). Essentially, the recommender periodically reorders the remaining items based on user feedback. The final recommendations can be directly compared to the initial list and ground-truth recommendations using metrics such as Recall@N and NDCG@N.

\textbf{Benefits of Simulated Environments.}\;\;
Simulated environments provide a safe training space for RL-based recommendation agents~\citep{putta2024agentqadvancedreasoning}, where they can learn from errors without inflicting them on real users. They enable modeling of long-term user feedback over a session or across sessions~\citep{zhao2023kuaisimcomprehensivesimulatorrecommender}, unlike conventional offline evaluations that focus on immediate responses, e.g., clicks. Further, LLM-driven simulators add interpretability by allowing simulated users to articulate their reasoning with thought traces, and give rich, natural language feedback, e.g. ``\emph{I don't like this recommendation because I've seen something similar}''. Simulators also help alleviate data-sharing concerns. Some simulators combine real item data with synthetic user populations. So one can test algorithms on synthetic yet behaviorally realistic users without exposing any personal user data. 

\vspace{-0.1in}
\section{User Simulation}
\label{sec:user-sim}

LLMs' strong reasoning and content understanding capabilities offer significant potential for simulating human behavior~\citep{Binz2025}. We simulate real user behavior by predicting their next action when presented with a list of recommended items. As illustrated in Figure~\ref{fig:example}, the simulated user's action space $\mathcal{A}$ for each item includes: (1) Click, (2) Comment, (3) Share, (4) Like, (5) Watch [specify duration in seconds], (6) Skip, and (7) Leave the session. If the user chooses to leave (action 7), they are prompted to reflect on their experience, specify reasons for dissatisfaction, and provide an instruction to the RecSys for improvement, or they may exit without further input. User decisions are influenced by both the current context $\mathcal{C}$ and past interactions $\mathcal{H}$. As shown in Figure~\ref{fig:example}, we consider contextual factors related to the user, such as temporal (time of day, seasonality), demographic (age, gender, location, etc.), behavioral (time spent, search queries, etc.), and social connections (group affiliations, etc.). 

A session begins when the user opens the app and ends when they exit. Each recommendation presents a list of $k$ items, selected from a candidate set $\mathcal{{I}}$ and shown in order. The agent and simulated user may interact over multiple turns within a session, e.g., the user requests more interesting content, and the agent updates the list. A reward signal is generated after each list is shown. Crucially, we aim to optimize for long-term user retention as the reward signal, i.e., maximizing session duration and minimizing the interval between sessions, which relates to Daily Active Users (DAU). Specifically, the simulated user goes through each recommendation one by one. For each item, the user takes three steps: 1. \textbf{Think it through}: \emph{What is your reasoning behind how you'll respond?} 2. \textbf{Take action}: \emph{Write down what you'll actually do.} 3. \textbf{Update your mindset}: \emph{How does this affect your current thinking?} Table~\ref{fig:output} shows a summary containing initial simulator results, including simulated user actions and reasoning for each item.

\vspace{-0.1in}
\subsection{Multimodal User Engagment Modeling}
\label{sec:item-metadata}

Effectively modeling user engagement history is essential for building simulators. 
A user may engage with multimodal items spanning text, audio, images, and videos. Processing them within a unified semantic space is nontrivial. Incorporating extended histories further requires the simulator to handle long contexts, with recent interactions likely carrying greater weight. For example, summarizing that ``\emph{in the past 7 days, the user watched 10 videos on deep-sea fishing, and commented on 5 friends' posts}'' can provide useful insights. Such compression can be achieved through statistical aggregation, semantic embeddings, or RAG techniques, all of which improve simulation efficiency. 
Below, we present three modeling alternatives for engagement history that leverage LLMs' powerful reasoning capabilities (Figure~\ref{fig:example}).

\begin{itemize}[topsep=3pt,itemsep=0pt,leftmargin=*]

\item \textbf{Text-based modeling}.\quad One approach is to represent all user and item data in textual form $\mathcal{T}$. User demographics can be specified in the system prompt, while each item’s metadata, such as publish time, genre, creator, content type, interest cluster, and a brief summary, together with the user’s interaction history, are provided as textual context. 
Moreover, when short videos are generated directly from user prompts, such as with Sora 2~\citep{sora2_openai_2025}, the prompts can be seamlessly incorporated into the item description.
This approach offers flexibility in determining which information to include in lifelong user behaviors, supports prompt compression for improved efficiency, and leverages the reasoning ability of LLMs to analyze user behavior and replicate user responses. 
Nonetheless, relying solely on textual descriptions may overlook certain nuances, including the mood of a video, sarcasm in content, or musical characteristics, which can limit the fidelity of preference modeling. 

\item \textbf{Multimodal modeling}.\quad The second option is to present each item with multimodal information $\mathcal{M}$, using a multimodal LLM (MLLM) as the backbone, such as Qwen3's Omni model~\citep{xu2025qwen2} or vision-language models (VLM), Gemini-2.5-Pro. In this setting, items are directly ingested by the model rather than represented through textual descriptions. Images are tokenized into patches, while short-form videos with their audio tracks can be jointly processed through frameworks that synchronize visual and acoustic signals. Demographic attributes and textual metadata can also be incorporated as auxiliary inputs. The model output may take multiple modalities, though we restrict it to text for predicting user behavior. Compared with text-only LLMs, multimodal models provide richer representational capacity for multimodal content, enabling more faithful input understanding. Nevertheless, their reasoning ability,  typically weaker, and the substantially larger token space of audio–visual inputs imposes significant context and efficiency challenges, making prompt compression particularly difficult.

\item \textbf{Semantic ID modeling}.\quad The third approach is to represent items using semantic IDs $\mathcal{S}$, where each item’s content, including video and affiliated multimodal features such as audio and textual tags, is encoded into a semantic ID~\citep{rajput2023recommender,deng2025onerec,zheng2025enhancing}. Items with similar semantics share similar IDs, with the sequence from the first to the last character encoding progressively finer-grained semantic information. 
During continuous pre-training, these semantic IDs are integrated into text descriptions so that their vector representations are learned jointly with the LLMs. 
At inference, user demographic information is provided in the prompt, while historical item engagements are represented using semantic IDs rather than textual descriptions or raw multimedia inputs. This approach leverages the strong reasoning capabilities of language models while providing a structured, compact representation of items. 
A limitation is that continuous pre-training is required to obtain a well-trained model, and periodic retraining is necessary to incorporate new content, such as newly released music albums. Unlike the first two approaches, no continual pre-training, supervised fine-tuning, or reinforcement learning is performed on the backbone LLMs or VLMs in this setting. 

\end{itemize}

\begin{table*}
\setlength{\tabcolsep}{4.5pt}
\renewcommand{\arraystretch}{1.15}
\centering
\textsf{
\begin{tabular}{llll}
\textbf{Video} & \textbf{Reasoning} & \textbf{Action Taken} & \textbf{Mindset Update}\\
\toprule
1. Lobster Fishing & Related to fishing interest & Watch (20s) & Open to more fishing content\\
2. UFC Fight Night & Strong interest, previous engagement & Watch (30s), Like & UFC is a top interest, expect more recommendations\\
3. Hairstyling & Unrelated, test new content & Skip & Not interested, less likely to engage\\
4. UFC Fight Night & Interest, but avoid over-engagement & Watch (15s), Skip & Selective with UFC, want variety\\
5. Hairstyling & Unrelated, already skipped & Skip & Not interested, expect fewer recommendations\\
\bottomrule
\end{tabular}}
\vspace{0.05in}
\caption{LLM uses their reasoning abilities to simulate user responses step-by-step. Using GPT-4.1, we set up a user profile: a 30-year-old male interested in deep sea fishing and has a history of liking UFC events. The RecSys serves him five short videos (see 1st column). For each, the simulator performs three tasks: ``\texttt{Reasoning}'' (2nd column), predicts the ``\texttt{Action Taken}'' (3rd column), and updates the ``\texttt{Mindset}'' (4th column). }
\label{fig:output}
\vspace{-0.25in}
\end{table*}

\textbf{Lifelong User Behavior Modeling}.\quad
In addition to effective multimodal user engagement modeling, it is essential to achieve robust \textit{lifelong user behavior modeling}. Users continuously interact with diverse items in recommender systems, resulting in ever-growing interaction histories. Over time, these histories can become effectively unbounded, making it computationally infeasible and costly to store and process all past behaviors. 
Precisely, we highlight two crucial principles built upon multimodal modeling: 

\begin{itemize}[leftmargin=*]
    \item \textbf{Dynamic memory modeling}.\quad Users may have infinite historical interactions $\mathcal{H}$. 
    However, not all behaviors contribute equally to reflecting the user’s current intention, and many may introduce noise. 
    It is therefore crucial to design mechanisms that can dynamically and intelligently store, filter, and retrieve user historical behaviors. 
    In RecoWorld, we define an \textit{engagement memory} $\mathcal{M}_u^t \subseteq \mathcal{H}_u$ at time $t$:
    \begin{equation}\small
    \mathcal{M}_u^t = \{ (a_k, i_k, \tau_k) \in \mathcal{H}_u \mid \alpha_k > \delta \},
    \end{equation}
    where $a_k$ denotes the action type (e.g., click, like) taken by the user at the $k$-th interaction, $i_k$ is the item interacted with, and $\tau_k$ is the timestamp of the interaction. $\alpha_k = h(a_k, i_k, \tau_k \mid \mathcal{C}_t)$ is the learned importance score conditioned on current context $\mathcal{C}_t$, and $\delta$ is a retention threshold.
    This enables dynamic filtering and retrieval of informative past behaviors while discarding noisy or outdated interactions. 
    Session-wise behaviors are also essential to model the user preference. Therefore, we organize the engagement memory at two levels: 
    \textit{Interaction-wise memory}:
    \begin{equation}
    \mathcal{M}_u^{t, \text{int}} = \{ (a_k, i_k) \}_{k=1}^t,
    \end{equation}
    which records fine-grained user actions $a_k$ and the corresponding items $i_k$. 
    2) \textit{Session-wise memory}: We group interactions into sessions $\mathcal{S}_u = \{s_1, s_2, \dots, s_m\}$, where each session $s_j$ is a trajectory of actions.
    A session summarization function $\phi(\cdot)$ produces latent representations:
    $
    \bm{s}_j = \phi(s_j) = \phi\big( \{(a_k, i_k) \in s_j\}\big),
    $
    and the session-wise memory is defined as:
    \begin{equation}
    \mathcal{M}_u^{t, \text{sess}} = \{ \bm{s}_1, \bm{s}_2, \dots, \bm{s}_m \},
    \end{equation}
    capturing trajectory-level signals such as mindset shifts and cross-session preference transitions.
    \item \textbf{Evolving preference modeling}.\quad
    User interests are not static but continuously evolve with changing contexts $\mathcal{C}_t$ and the information they are exposed to through recommendations. 
    We denote the user’s latent preference state at time $t$ as $\bm{z}_t$. Its evolution can be modeled as:
    \begin{equation}
    \bm{z}_t = g\big(\bm{z}_{t-1}, \mathcal{M}_u^t, \mathcal{C}_t \big),
    \end{equation}
    where $\mathcal{M}_u^t$ is the engagement memory (interaction-wise or session-wise), $\mathcal{C}_t$ is the current context, and $g(\cdot)$ is a temporal update function (e.g., recurrent, attention-based, or diffusion-style).
    To capture fine-grained preference shifts, we further decompose the update into a \textit{mindset update}:
    \begin{equation}
    \bm{z}_t = \bm{z}_{t-1} + \Delta\bm{m}_t, \quad
    \Delta\bm{m}_t = f\big(\mathcal{M}_u^t, \mathcal{C}_t, r_t, \Delta_t \big),
    \end{equation}
    where $\Delta\bm{m}_t$ is the mindset adjustment derived from the current memory $\mathcal{M}_u^t$, context $\mathcal{C}_t$, reasoning trace $r_t$, and any correction applied by the system to the user's preference state $\Delta_t$.    
    This formulation ensures that the simulator captures temporal dynamics, long-term dependencies, and potential preference shifts, such that the generated behaviors 
    remain personalized and timely throughout the interaction lifecycle (see Figure~\ref{fig:output}).
\end{itemize}

\vspace{-0.1in}
\subsection{Candidate Items}
\label{sec:item-metadata}

We use real-world content whenever possible (e.g., engagement history and recommended items) to ensure realistic user behavior modeling. Using both simulated content and simulated users can be challenging, as combining these approximations may affect the reliability of behavioral insights. 
Specifically, each item is presented by its content (e.g., short-form video) along with its metadata and text description. 
The metadata includes details such as the timestamp, creator information, engagement metrics (e.g., likes, shares, comments), and content category. A user may be recommended various types of content, such as: (a) \textbf{short-form videos}, (b) \textbf{text posts}, (c) \textbf{mixed media} (posts, videos, and images), and (d) \textbf{friend suggestions}. 
Multiple representations can be used for content understanding: text descriptions provide a summary of the image or video's content; audio/visual encoders extract multimodal features from the media; and semantic IDs capture high-level semantic categories. The simulator may also leverage metadata tags, such as hashtags or timestamps, to better mimic realistic user behavior.

\begin{figure*}
\centering
\includegraphics[width=0.78\textwidth]{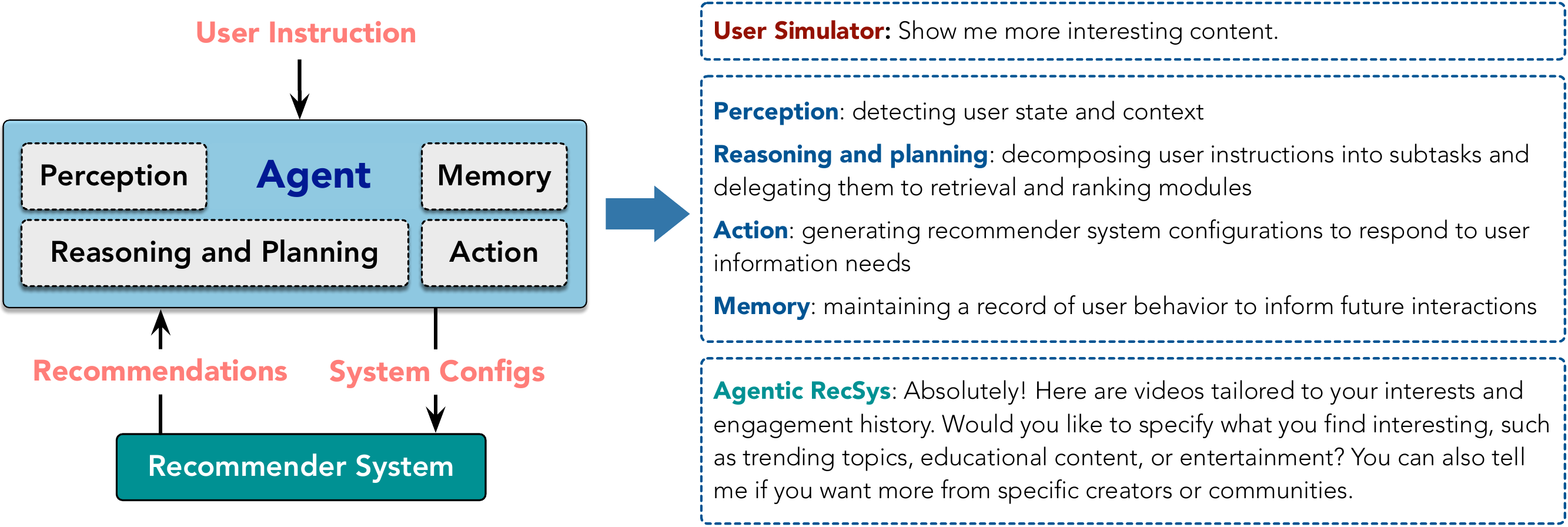}
\vspace{-0.1in}
\caption{An instruction-following recommender can be powered by an autonomous agent.}
\label{fig:inforec}
\vspace{-0.1in}
\end{figure*}

\section{Instruction-Following RecSys}
\label{sec:info_rec}

Once users get used to giving instructions to LLMs, they tend to do the same with every system they use~\citep{zhou2023instructionfollowingevaluationlargelanguage,brooks2023instructpix2pixlearningfollowimage}. For recommender systems, this means developing new agentic capabilities that allow them to respond to user commands and adjust their behavior accordingly. Users might ask to see more or less of the content that interests them, request content from specific sources such as particular friends or topics, or ask questions about the content they are seeing in their screen. This applies to various media forms, ranging from text, photos, videos, multimedia carousels, to products and even social connections.

An instruction-following recommender could be built around an \textbf{autonomous agent} with four core capabilities (see Figure~\ref{fig:inforec}): (a) \emph{perception}: detecting user state and context; (b) \emph{reasoning and planning}: decomposing user instructions into subtasks and delegating them to retrieval and ranking modules; (c) \emph{action (tool use)}: delivering updated recommendations and responding to user information needs; and (d) \emph{memory}: maintaining a record of user behavior to inform future interactions.

In the context of recommender systems, the interaction between a user and the system can be modeled as a Markov Decision Process. The task involves a series of interleaved states and actions $(s_0, a_0, s_1, \ldots, a_{T-1}, s_T)$, beginning at an initial state $s_0$ and finally reach the goal state $s_T$. Here, the environment state $s_t$ represents the user's mindset, which evolves as the user interacts with recommended items. The state transition is modeled as $p_{\theta}(s_{t+1} | s_t, a_t)$. It captures how the user's mindset changes after the recommender system presents a new set of items, and this process can be simulated using a LLM-based user simulator, with parameters $\theta$. At each step, the agentic recommender system selects an action $a_t \sim p_{\phi}(a|s_t)$, yielding a new list of recommendations by considering the user's explicit or implicit instruction provided by $s_t$.

The recommender model's parameters $\phi$ could be related to its perception, memory, planning, tool use modules or the architecture itself in the case of generative recommenders. These parameters are updated iteratively to improve the policy. The ultimate goal is to develop a policy, denoted as $a_t^* = p_{\phi}(a|s_t)$, which focuses on identifying the optimal recommendations $a_t^*$ given the current state $s_t$, that maximizes the total expected reward over time. A reward function $\mathcal{R}$ assigns a scalar reward $\mathcal{R}_{\eta}(r_t | s_t, a_t)$ after an action is taken from a given state. We can perform multiple rollouts to estimate cumulative rewards, using on-policy learning methods such as PPO with asynchronous rewards or off-policy learning such as DPO. The reward function may incorporate metrics such as total session time, number of clicks, self-critique scores, or a combination of factors that measure the system's ability to follow user instructions and maximize engagement. This approach provides more robust feedback compared to action-level rewards.

\textbf{Voice-Driven Feedback.}\;\; There is increasing business interest in developing agentic recommender systems. Companies have introduced easy-to-access voice entry points that allow users to adjust the content they see in their feed, friend graph, and across various product surfaces. E.g., a user might say ``\emph{I want more content about F1,}'' ``\emph{Show me more Reels like this,}'' ``\emph{Show me what’s new with her,}'' ``\emph{Show me content from my high school classmates,}'' ``\emph{Show me the latest posts from my family,}'' ``\emph{Show me more `interesting’ content}'' and ``\emph{Show me less political content.}'' Such voice-based user instructions provide valuable feedback signals that agentic recommender systems can learn from.

\textbf{Instruction Alignment.}\quad\;\; User instructions can span a broad spectrum, so it is necessary to align them accurately with the underlying user intents~\citep{Wang_2025}. For example, a user requesting more content on ``\emph{UFC Fight Night}'' may have different motivations: they might be looking for a fun local event to enjoy with friends (a short-term, situational interest where other local activities could also appeal), or they could be exploring combat sports, possibly with a long-term interest of training for competitions (in which case Brazilian Jiu-Jitsu would be relevant). Users may provide vague or open-ended feedback, such as ``\emph{I’m bored,}'' which poses a greater challenge than responding to specific requests such as ``\emph{I’d like to see long skirts in light blue for around \$200}''~\citep{tang2025interactiverecommendationagentactive}. While lexical or semantic matching can typically align user queries with relevant content, ambiguous instructions require more sophisticated handling. We envision a reasoning-intensive retrieval model powered by LLMs to infer user intents~\citep{xing2026reg4recreasoningenhancedgenerativemodel}, refine ambiguous queries, and enhance the system's ability to understand user instructions.

\textbf{Instruction-Following (InFoRec) vs. Conversational Recommender System (CRS)}\;\;
The key difference is that InFoRec respond directly to user commands, while CRS takes a more proactive approach, suggesting items during the conversation. For example, in a CRS chat: User: ``\emph{Hi, what's the date today?}" CRS: ``\emph{Today is September 16, 2025. By the way, have you heard of Brad Pitt's new movie `F1'? It's super popular right now}''~\citep{wen2025usbreceffectiveframeworkimproving,zhao2025exploringimpactpersonalitytraits}. Previous research has developed various persuasion strategies, such as authority, emotional appeal, social proof, to recommend items tailored to users’ personality traits. In contrast, InFoRec focuses on precisely following user instructions. It refines recommendations by adding, removing, or reordering items among the top results. Think of InFoRec as a tool that is triggered during the conversation: by utilizing the entire conversation and engagement history, InFoRec dynamically adjusts the recommendations shown to the user.

\vspace{-0.1in}
\section{Multi-Agent Simulators}
\label{sec:multi-agent}

Recent advances in multi-agent simulations have substantially enhanced our ability to model the dynamics of individuals making decisions and interacting over time~\citep{park2023generativeagentsinteractivesimulacra,park2024generativeagentsimulations1000,dubois2024alpacafarmsimulationframeworkmethods,gao2024agentscopeflexiblerobustmultiagent,yang2025oasisopenagentsocial,tang2025gensimgeneralsocialsimulation,louie2025llmsimulatedpracticefeedbackupskill,qian2025userbenchinteractivegymenvironment,anthis2025llmsocialsimulationspromising}. 
Particularly, \citet{anthis2025llmsocialsimulationspromising} highlight the potential of LLM-based social simulations. They identify five key tractable challenges that pave the way for innovations in this space. AgentTorch~\citep{chopra2024limitsagencyagentbasedmodels} uses the COVID-19 pandemic to test whether LLM agents can replicate population-level behavioral statistics. It further examines the trade-off between simulation scale and individual agency. OASIS~\citep{yang2025oasisopenagentsocial} simulates social media platforms, e.g., X and Reddit, to study information diffusion, group polarization, and herd behavior. AgentSociety~\citep{piao2025agentsocietylargescalesimulationllmdriven} provides a testbed for computational social experiments, examining polarization, the spread of inflammatory content, universal basic income, and the impact of shocks such as hurricanes. These simulators vary in scale, from thousands of agents to millions (e.g., OASIS, AgentScope).

Consider an environment consisting of $N$ simulated users. We denote by $u_i(t)$ the state of user $i$ at simulation time $t$, which includes a set of user attributes such as age, location, interests, and the set of recommendations with which they have interacted, among others. As the simulation progresses, user $i$ updates their state $u_i(t)$ by interacting with their neighbors $\mathcal N_i(t)$ and the environment $e(t)$, both of which may also evolve over time. The neighborhood of a user can be defined by a social network graph, similarity metrics, or other criteria. We denote by $s_{ij}(t)$ the information that user $i$ receives from their interaction with user $j$. For example, $s_{ij}(t)$ may represent a shared post or a message exchanged between users. Following~\citet{chopra2024limitsagencyagentbasedmodels}, the user's update rule is specified in Eq. (\ref{eq:user_update}), and the environment evolves in response to users' actions and states, as described by Eq. (\ref{eq:env_update}). Both $f$ and $g$ are governed by the recommender system dynamics. The strength of user-to-user influence is specified by environment parameters. 
\begin{align*}
u_i(t+1) &= f\left(u_i(t), \cup_{j \in \mathcal N(i)} s_{ij}(t), e(t) \right).
\numberthis\label{eq:user_update}\\
e(t+1) &= g\left(\{u_i(t+1)\}_{i=1}^N, e(t) \right).
\numberthis\label{eq:env_update}
\end{align*}

In recommender systems, these tools show great promise for creator support. They may provide projections to optimize content planning and distribution, which helps creators grow their audience. When a creator compiles a new video, they may want to ``test the waters'' by exposing this content to a group of simulated users to gauge potential audience reactions. These simulated users can represent the creator’s existing followers, a general population sample, or a targeted demographic (e.g., teens in the US and Canada). Creators might use this simulation to experiment with different publishing strategies, such as adjusting frequency to avoid audience fatigue, diversifying content to attract new viewers, or even publishing controversial material to boost engagement. Without a simulator, the outcomes of these strategies are often unpredictable and, if mishandled, could have negative consequences. By leveraging a network of interconnected simulated users, we can aggregate collective feedback and track predicted user responses at multiple intervals: 6 hours, 12 hours, 24 hours, 2 days, and 1 week post-exposure. Additionally, the simulation can be reset and rerun to measure variability and improve confidence in the predictions. This approach provides an early estimate of how the content might perform in the real world.

\section{Related Work}
\label{sec:related}

\textbf{Evaluation of Recommender Systems.}\quad
Recommender systems have seen tremendous progress in recent years~\citep{wang2024llmsuserexplorationlargescale,hou2025actionpiececontextuallytokenizingaction,zhu2025longtermclockfinegrainedtime,yoon2025omuletorchestratingmultipletools,xing2025reg4recreasoningenhancedgenerativemodel,zhang2025notellm2multimodallargerepresentation,chen2025pinfmfoundationmodeluser,deng2025onerecunifyingretrieverank,he2025plumadaptingpretrainedlanguage,zhou2026openonerectechnicalreport}, with generative recommenders, e.g. Meta's HSTU and Kuaishou’s OneRec, and emerging agentic recommenders holding promise for delivering highly personalized content. For example, HSTU achieves up to 65\% better ranking metrics than baselines, with a 1.5-trillion-parameter version boosting online A/B metrics by 12.4\%~\citep{zhai2024actionsspeaklouderwords}. Similarly, OneRec replaces the traditional retrieve-then-rank pipeline with an end-to-end generative model and was deployed to millions of users, yielding a +1.6\% increase in watch-time. A critical challenge comes with these more powerful models is evaluation. Traditional offline metrics often prove insufficient: promising ideas are dropped due to unconvincing offline results, and improvements that look good offline do not always translate to real user gains. Yet running A/B tests is costly and time-consuming, and it is impractical (and risky) to expose users to every experiment. This has led to growing interest in large-scale simulated environments for recommender evaluation.

\textbf{Simulated Environments for Training Agentic Systems.}
Simulated environments have been pivotal in training autonomous agents through RL, ranging from physics sandboxes such as MuJoCo~\citep{zakka2025mujocoplayground}, to games in the Arcade Learning Environment~\citep{farebrother2024calecontinuousarcadelearning}, and to computer-use tasks such as WebArena~\citep{zhou2024webarenarealisticwebenvironment}. World models, which are learned simulators of environment dynamics, have also played a significant role~\citep{fung2025embodiedaiagentsmodeling}. In RecSys, one notable effort is KuaiSim~\citep{zhao2023kuaisimcomprehensivesimulatorrecommender}, which provides a rich user environment with multi-behavior feedback and cross-session interactions. This enables training RL-based recommenders on sessions and retention scenarios~\citep{zhang2025darlrdualagentofflinereinforcement}. Another example is AgentRecBench~\citep{shang2025agentrecbenchbenchmarkingllmagentbased}, which constructs a simulation environment by processing multi-domain datasets (e.g., Yelp, Goodreads, Amazon) and exposing standardized ``world'' APIs for an agent to query and act. These environments provide a sandbox for agents to plan, explore, and learn complex behaviors.

\vspace{0.05in}
\textbf{User Simulation with LLMs.}\;
LLM-empowered user simulator has shown promising results in imitating real-world user behaviors across different domains, such as social simulation~\citep{gurcan2023llm}, legal judgement~\citep{jiang2025agentsbench}, and recommender systems~\citep{bougie2025simusersimulatinguserbehavior,parkchoongwon2025,peng2025survey,chen2025vragentr1boostingvideorecommendation,liu2026diagnosticguideddynamicprofileoptimization}. 
In RecSys, LLM agents are used to generate simulated user feedback for model evaluation~\citep{yoon2024evaluating,wei2025mirroringusersbuildingpreferencealigned,Chen_2025,zhang2026exploringrecommenderevaluationmultimodal} and optimization~\citep{cai2025agentic,liu2025diagnosticguideddynamicprofileoptimization}. 
Existing works can be mainly categorized into two groups: 
1) Non-conversational user simulators~\citep{zhang2025llm}: These LLM agents give realistic feedback (e.g., like, dislike) on recommendations, as exemplified by Agent4Rec~\citep{zhang2024generative} and AgentCF~\citep{zhang2024agentcf}. Typically, non-conversational user simulator cannot proactively give instructions to recommender systems or interact with environments (e.g., other LLM agents). 
Another line of work lies in 2) Conversational user simulators~\citep{guo2024knowledge,yoon2024evaluating}: These agents engage in dialogue with recommender models and other agents (e.g., RecAgent~\citep{wang2025user}), often incorporating self-reflection mechanisms to more accurately model user intent. 
To support the development of agentic RecSys, we borrow the idea from conversational user simulators: they can issue context-aware instructions at appropriate moments. E.g., indicating a desire to shift away from prior preferences due to fatigue or boredom with previous recommendations.

\vspace{-0.1in}
\section{Industry Use and Conclusion}
\label{sec:conclusion}

Our \textsc{RecoWorld} introduces a novel environment for instruction-driven interactions between users and agentic recommender systems. Internally released in September 2025, this work establishes the foundation for instruction-following recommenders. In this paper, we prioritize presenting a comprehensive framework for building simulated environments. We note that \textsc{RecoWorld} has already received significant interest from leading industrial stakeholders, including Google and Kuaishou RecSys teams, suggesting its potential impact on real-world applications.


\balance
\bibliographystyle{ACM-Reference-Format}
\bibliography{custom}

@article{zheng2025enhancing,
  title={Enhancing Embedding Representation Stability in Recommendation Systems with Semantic ID},
  author={Zheng, Carolina and Huang, Minhui and Pedchenko, Dmitrii and Rangadurai, Kaushik and Wang, Siyu and Nahum, Gaby and Lei, Jie and Yang, Yang and Liu, Tao and Luo, Zutian and others},
  journal={arXiv preprint arXiv:2504.02137},
  year={2025}
}

@article{deng2025onerec,
  title={Onerec: Unifying retrieve and rank with generative recommender and iterative preference alignment},
  author={Deng, Jiaxin and Wang, Shiyao and Cai, Kuo and Ren, Lejian and Hu, Qigen and Ding, Weifeng and Luo, Qiang and Zhou, Guorui},
  journal={arXiv preprint arXiv:2502.18965},
  year={2025}
}

@article{rajput2023recommender,
  title={Recommender systems with generative retrieval},
  author={Rajput, Shashank and Mehta, Nikhil and Singh, Anima and Hulikal Keshavan, Raghunandan and Vu, Trung and Heldt, Lukasz and Hong, Lichan and Tay, Yi and Tran, Vinh and Samost, Jonah and others},
  journal={Advances in Neural Information Processing Systems},
  volume={36},
  pages={10299--10315},
  year={2023}
}

@article{peng2025survey,
  title={A survey on llm-powered agents for recommender systems},
  author={Peng, Qiyao and Liu, Hongtao and Huang, Hua and Yang, Qing and Shao, Minglai},
  journal={arXiv preprint arXiv:2502.10050},
  year={2025}
}

@article{jiang2025agentsbench,
  title={AgentsBench: A Multi-Agent LLM Simulation Framework for Legal Judgment Prediction},
  author={Jiang, Cong and Yang, Xiaolei},
  journal={Systems},
  volume={13},
  number={8},
  pages={641},
  year={2025},
  publisher={MDPI}
}

@article{gurcan2023llm,
  title={LLM-Augmented Agent-Based Modelling for Social Simulations: Challenges and Opportunities},
  author={G{\"u}rcan, Onder},
  journal={HHAI 2024: Hybrid Human AI Systems for the Social Good},
  pages={134},
  year={2023}
}

@inproceedings{zhang2025llm,
  title={Llm-powered user simulator for recommender system},
  author={Zhang, Zijian and Liu, Shuchang and Liu, Ziru and Zhong, Rui and Cai, Qingpeng and Zhao, Xiangyu and Zhang, Chunxu and Liu, Qidong and Jiang, Peng},
  booktitle={Proceedings of the AAAI Conference on Artificial Intelligence},
  volume={39},
  number={12},
  pages={13339--13347},
  year={2025}
}

@article{guo2024knowledge,
  title={Knowledge graph enhanced language agents for recommendation},
  author={Guo, Taicheng and Liu, Chaochun and Wang, Hai and Mannam, Varun and Wang, Fang and Chen, Xin and Zhang, Xiangliang and Reddy, Chandan K},
  journal={arXiv preprint arXiv:2410.19627},
  year={2024}
}

@article{yoon2024evaluating,
  title={Evaluating large language models as generative user simulators for conversational recommendation},
  author={Yoon, Se-eun and He, Zhankui and Echterhoff, Jessica Maria and McAuley, Julian},
  journal={arXiv preprint arXiv:2403.09738},
  year={2024}
}

@article{wang2025user,
  title={User behavior simulation with large language model-based agents},
  author={Wang, Lei and Zhang, Jingsen and Yang, Hao and Chen, Zhi-Yuan and Tang, Jiakai and Zhang, Zeyu and Chen, Xu and Lin, Yankai and Sun, Hao and Song, Ruihua and others},
  journal={ACM Transactions on Information Systems},
  volume={43},
  number={2},
  pages={1--37},
  year={2025},
  publisher={ACM New York, NY}
}

@inproceedings{zhang2024agentcf,
  title={Agentcf: Collaborative learning with autonomous language agents for recommender systems},
  author={Zhang, Junjie and Hou, Yupeng and Xie, Ruobing and Sun, Wenqi and McAuley, Julian and Zhao, Wayne Xin and Lin, Leyu and Wen, Ji-Rong},
  booktitle={Proceedings of the ACM Web Conference 2024},
  pages={3679--3689},
  year={2024}
}

@inproceedings{zhang2024generative,
  title={On generative agents in recommendation},
  author={Zhang, An and Chen, Yuxin and Sheng, Leheng and Wang, Xiang and Chua, Tat-Seng},
  booktitle={Proceedings of the 47th international ACM SIGIR conference on research and development in Information Retrieval},
  pages={1807--1817},
  year={2024}
}

@inproceedings{cai2025agentic,
  title={Agentic feedback loop modeling improves recommendation and user simulation},
  author={Cai, Shihao and Zhang, Jizhi and Bao, Keqin and Gao, Chongming and Wang, Qifan and Feng, Fuli and He, Xiangnan},
  booktitle={Proceedings of the 48th International ACM SIGIR conference on Research and Development in Information Retrieval},
  pages={2235--2244},
  year={2025}
}

@misc{xu2025qwen2,
  title={Qwen2. 5-omni technical report},
  author={Xu, Jin and Guo, Zhifang and He, Jinzheng and Hu, Hangrui and He, Ting and Bai, Shuai and Chen, Keqin and Wang, Jialin and Fan, Yang and Dang, Kai and others},
  journal={arXiv preprint arXiv:2503.20215},
  year={2025}
}

@misc{corecco2024suberrlenvironmentsimulated,
      title={SUBER: An RL Environment with Simulated Human Behavior for Recommender Systems}, 
      author={Nathan Corecco and Giorgio Piatti and Luca A. Lanzendörfer and Flint Xiaofeng Fan and Roger Wattenhofer},
      year={2024},
      eprint={2406.01631},
      archivePrefix={arXiv},
      primaryClass={cs.IR},
      url={https://arxiv.org/abs/2406.01631}, 
}

@misc{liu2025diagnosticguideddynamicprofileoptimization,
      title={Diagnostic-Guided Dynamic Profile Optimization for LLM-based User Simulators in Sequential Recommendation}, 
      author={Hongyang Liu and Zhu Sun and Tianjun Wei and Yan Wang and Jiajie Zhu and Xinghua Qu},
      year={2025},
      eprint={2508.12645},
      archivePrefix={arXiv},
      primaryClass={cs.IR},
      url={https://arxiv.org/abs/2508.12645}, 
}

@misc{wei2025mirroringusersbuildingpreferencealigned,
      title={Mirroring Users: Towards Building Preference-aligned User Simulator with User Feedback in Recommendation}, 
      author={Tianjun Wei and Huizhong Guo and Yingpeng Du and Zhu Sun and Chen Huang and Dongxia Wang and Jie Zhang},
      year={2025},
      eprint={2508.18142},
      archivePrefix={arXiv},
      primaryClass={cs.HC},
      url={https://arxiv.org/abs/2508.18142}, 
}

@misc{xing2025reg4recreasoningenhancedgenerativemodel,
      title={REG4Rec: Reasoning-Enhanced Generative Model for Large-Scale Recommendation Systems}, 
      author={Haibo Xing and Hao Deng and Yucheng Mao and Jinxin Hu and Yi Xu and Hao Zhang and Jiahao Wang and Shizhun Wang and Yu Zhang and Xiaoyi Zeng and Jing Zhang},
      year={2025},
      eprint={2508.15308},
      archivePrefix={arXiv},
      primaryClass={cs.IR},
      url={https://arxiv.org/abs/2508.15308}, 
}

@inproceedings{Chen_2025, series={WWW ’25},
   title={RecUserSim: A Realistic and Diverse User Simulator for Evaluating Conversational Recommender Systems},
   url={http://dx.doi.org/10.1145/3701716.3715258},
   DOI={10.1145/3701716.3715258},
   booktitle={Companion Proceedings of the ACM on Web Conference 2025},
   publisher={ACM},
   author={Chen, Luyu and Dai, Quanyu and Zhang, Zeyu and Feng, Xueyang and Zhang, Mingyu and Tang, Pengcheng and Chen, Xu and Zhu, Yue and Dong, Zhenhua},
   year={2025},
   month=may, pages={133–142},
   collection={WWW ’25} }

@misc{tang2025gensimgeneralsocialsimulation,
      title={GenSim: A General Social Simulation Platform with Large Language Model based Agents}, 
      author={Jiakai Tang and Heyang Gao and Xuchen Pan and Lei Wang and Haoran Tan and Dawei Gao and Yushuo Chen and Xu Chen and Yankai Lin and Yaliang Li and Bolin Ding and Jingren Zhou and Jun Wang and Ji-Rong Wen},
      year={2025},
      eprint={2410.04360},
      archivePrefix={arXiv},
      primaryClass={cs.MA},
      url={https://arxiv.org/abs/2410.04360}, 
}

@misc{tang2025interactiverecommendationagentactive,
      title={Interactive Recommendation Agent with Active User Commands}, 
      author={Jiakai Tang and Yujie Luo and Xunke Xi and Fei Sun and Xueyang Feng and Sunhao Dai and Chao Yi and Dian Chen and Zhujin Gao and Yang Li and Xu Chen and Wen Chen and Jian Wu and Yuning Jiang and Bo Zheng},
      year={2025},
      eprint={2509.21317},
      archivePrefix={arXiv},
      primaryClass={cs.IR},
      url={https://arxiv.org/abs/2509.21317}, 
}

@misc{zhao2025exploringimpactpersonalitytraits,
      title={Exploring the Impact of Personality Traits on Conversational Recommender Systems: A Simulation with Large Language Models}, 
      author={Xiaoyan Zhao and Yang Deng and Wenjie Wang and Hongzhan lin and Hong Cheng and Rui Zhang and See-Kiong Ng and Tat-Seng Chua},
      year={2025},
      eprint={2504.12313},
      archivePrefix={arXiv},
      primaryClass={cs.CL},
      url={https://arxiv.org/abs/2504.12313}, 
}

@misc{qian2025userbenchinteractivegymenvironment,
      title={UserBench: An Interactive Gym Environment for User-Centric Agents}, 
      author={Cheng Qian and Zuxin Liu and Akshara Prabhakar and Zhiwei Liu and Jianguo Zhang and Haolin Chen and Heng Ji and Weiran Yao and Shelby Heinecke and Silvio Savarese and Caiming Xiong and Huan Wang},
      year={2025},
      eprint={2507.22034},
      archivePrefix={arXiv},
      primaryClass={cs.AI},
      url={https://arxiv.org/abs/2507.22034}, 
}

@misc{gao2024agentscopeflexiblerobustmultiagent,
      title={AgentScope: A Flexible yet Robust Multi-Agent Platform}, 
      author={Dawei Gao and Zitao Li and Xuchen Pan and Weirui Kuang and Zhijian Ma and Bingchen Qian and Fei Wei and Wenhao Zhang and Yuexiang Xie and Daoyuan Chen and Liuyi Yao and Hongyi Peng and Zeyu Zhang and Lin Zhu and Chen Cheng and Hongzhu Shi and Yaliang Li and Bolin Ding and Jingren Zhou},
      year={2024},
      eprint={2402.14034},
      archivePrefix={arXiv},
      primaryClass={cs.MA},
      url={https://arxiv.org/abs/2402.14034}, 
}

@misc{putta2024agentqadvancedreasoning,
      title={Agent Q: Advanced Reasoning and Learning for Autonomous AI Agents}, 
      author={Pranav Putta and Edmund Mills and Naman Garg and Sumeet Motwani and Chelsea Finn and Divyansh Garg and Rafael Rafailov},
      year={2024},
      eprint={2408.07199},
      archivePrefix={arXiv},
      primaryClass={cs.AI},
      url={https://arxiv.org/abs/2408.07199}, 
}

@techreport{anthropic2025claude,
  title = {System Card: Claude Opus 4 \& Claude Sonnet 4},
  author = {{Anthropic}},
  year = {2025},
  month = may,
  url = {https://www-cdn.anthropic.com/4263b940cabb546aa0e3283f35b686f4f3b2ff47.pdf},
  note = {Accessed: 2025-08-09}
}

@misc{fung2025embodiedaiagentsmodeling,
      title={Embodied AI Agents: Modeling the World}, 
      author={Pascale Fung and Yoram Bachrach and Asli Celikyilmaz and Kamalika Chaudhuri and Delong Chen and Willy Chung and Emmanuel Dupoux and Hongyu Gong and Hervé Jégou and Alessandro Lazaric and Arjun Majumdar and Andrea Madotto and Franziska Meier and Florian Metze and Louis-Philippe Morency and Théo Moutakanni and Juan Pino and Basile Terver and Joseph Tighe and Paden Tomasello and Jitendra Malik},
      year={2025},
      eprint={2506.22355},
      archivePrefix={arXiv},
      primaryClass={cs.AI},
      url={https://arxiv.org/abs/2506.22355}, 
}

@misc{zhang2025darlrdualagentofflinereinforcement,
      title={DARLR: Dual-Agent Offline Reinforcement Learning for Recommender Systems with Dynamic Reward}, 
      author={Yi Zhang and Ruihong Qiu and Xuwei Xu and Jiajun Liu and Sen Wang},
      year={2025},
      eprint={2505.07257},
      archivePrefix={arXiv},
      primaryClass={cs.IR},
      url={https://arxiv.org/abs/2505.07257}, 
}

@misc{brooks2023instructpix2pixlearningfollowimage,
      title={InstructPix2Pix: Learning to Follow Image Editing Instructions}, 
      author={Tim Brooks and Aleksander Holynski and Alexei A. Efros},
      year={2023},
      eprint={2211.09800},
      archivePrefix={arXiv},
      primaryClass={cs.CV},
      url={https://arxiv.org/abs/2211.09800}, 
}

@misc{sora2_openai_2025,
  title        = {Sora 2 is here},
  author       = {{OpenAI}},
  year         = {2025},
  howpublished = {\url{https://openai.com/index/sora-2/}},
  note         = {Accessed: 2025-10-06}
}

@misc{zhou2023instructionfollowingevaluationlargelanguage,
      title={Instruction-Following Evaluation for Large Language Models}, 
      author={Jeffrey Zhou and Tianjian Lu and Swaroop Mishra and Siddhartha Brahma and Sujoy Basu and Yi Luan and Denny Zhou and Le Hou},
      year={2023},
      eprint={2311.07911},
      archivePrefix={arXiv},
      primaryClass={cs.CL},
      url={https://arxiv.org/abs/2311.07911}, 
}

@misc{shang2025agentrecbenchbenchmarkingllmagentbased,
      title={AgentRecBench: Benchmarking LLM Agent-based Personalized Recommender Systems}, 
      author={Yu Shang and Peijie Liu and Yuwei Yan and Zijing Wu and Leheng Sheng and Yuanqing Yu and Chumeng Jiang and An Zhang and Fengli Xu and Yu Wang and Min Zhang and Yong Li},
      year={2025},
      eprint={2505.19623},
      archivePrefix={arXiv},
      primaryClass={cs.IR},
      url={https://arxiv.org/abs/2505.19623}, 
}

@misc{louie2025llmsimulatedpracticefeedbackupskill,
      title={Can LLM-Simulated Practice and Feedback Upskill Human Counselors? A Randomized Study with 90+ Novice Counselors}, 
      author={Ryan Louie and Ifdita Hasan Orney and Juan Pablo Pacheco and Raj Sanjay Shah and Emma Brunskill and Diyi Yang},
      year={2025},
      eprint={2505.02428},
      archivePrefix={arXiv},
      primaryClass={cs.HC},
      url={https://arxiv.org/abs/2505.02428}, 
}

@misc{park2024generativeagentsimulations1000,
      title={Generative Agent Simulations of 1,000 People}, 
      author={Joon Sung Park and Carolyn Q. Zou and Aaron Shaw and Benjamin Mako Hill and Carrie Cai and Meredith Ringel Morris and Robb Willer and Percy Liang and Michael S. Bernstein},
      year={2024},
      eprint={2411.10109},
      archivePrefix={arXiv},
      primaryClass={cs.AI},
      url={https://arxiv.org/abs/2411.10109}, 
}

@misc{anthis2025llmsocialsimulationspromising,
      title={LLM Social Simulations Are a Promising Research Method}, 
      author={Jacy Reese Anthis and Ryan Liu and Sean M. Richardson and Austin C. Kozlowski and Bernard Koch and James Evans and Erik Brynjolfsson and Michael Bernstein},
      year={2025},
      eprint={2504.02234},
      archivePrefix={arXiv},
      primaryClass={cs.HC},
      url={https://arxiv.org/abs/2504.02234}, 
}

@misc{yang2025oasisopenagentsocial,
      title={OASIS: Open Agent Social Interaction Simulations with One Million Agents}, 
      author={Ziyi Yang and Zaibin Zhang and Zirui Zheng and Yuxian Jiang and Ziyue Gan and Zhiyu Wang and Zijian Ling and Jinsong Chen and Martz Ma and Bowen Dong and Prateek Gupta and Shuyue Hu and Zhenfei Yin and Guohao Li and Xu Jia and Lijun Wang and Bernard Ghanem and Huchuan Lu and Chaochao Lu and Wanli Ouyang and Yu Qiao and Philip Torr and Jing Shao},
      year={2025},
      eprint={2411.11581},
      archivePrefix={arXiv},
      primaryClass={cs.CL},
      url={https://arxiv.org/abs/2411.11581}, 
}

@misc{dubois2024alpacafarmsimulationframeworkmethods,
      title={AlpacaFarm: A Simulation Framework for Methods that Learn from Human Feedback}, 
      author={Yann Dubois and Xuechen Li and Rohan Taori and Tianyi Zhang and Ishaan Gulrajani and Jimmy Ba and Carlos Guestrin and Percy Liang and Tatsunori B. Hashimoto},
      year={2024},
      eprint={2305.14387},
      archivePrefix={arXiv},
      primaryClass={cs.LG},
      url={https://arxiv.org/abs/2305.14387}, 
}

@misc{maragheh2025futureagenticdefinitionsperspectives,
      title={The Future is Agentic: Definitions, Perspectives, and Open Challenges of Multi-Agent Recommender Systems}, 
      author={Reza Yousefi Maragheh and Yashar Deldjoo},
      year={2025},
      eprint={2507.02097},
      archivePrefix={arXiv},
      primaryClass={cs.IR},
      url={https://arxiv.org/abs/2507.02097}, 
}

@misc{brown2024largelanguagemonkeysscaling,
      title={Large Language Monkeys: Scaling Inference Compute with Repeated Sampling}, 
      author={Bradley Brown and Jordan Juravsky and Ryan Ehrlich and Ronald Clark and Quoc V. Le and Christopher Ré and Azalia Mirhoseini},
      year={2024},
      eprint={2407.21787},
      archivePrefix={arXiv},
      primaryClass={cs.LG},
      url={https://arxiv.org/abs/2407.21787}, 
}

@misc{zhao2023kuaisimcomprehensivesimulatorrecommender,
      title={KuaiSim: A Comprehensive Simulator for Recommender Systems}, 
      author={Kesen Zhao and Shuchang Liu and Qingpeng Cai and Xiangyu Zhao and Ziru Liu and Dong Zheng and Peng Jiang and Kun Gai},
      year={2023},
      eprint={2309.12645},
      archivePrefix={arXiv},
      primaryClass={cs.IR},
      url={https://arxiv.org/abs/2309.12645}, 
}

@misc{farebrother2024calecontinuousarcadelearning,
      title={CALE: Continuous Arcade Learning Environment}, 
      author={Jesse Farebrother and Pablo Samuel Castro},
      year={2024},
      eprint={2410.23810},
      archivePrefix={arXiv},
      primaryClass={cs.LG},
      url={https://arxiv.org/abs/2410.23810}, 
}

@misc{zakka2025mujocoplayground,
      title={MuJoCo Playground}, 
      author={Kevin Zakka and Baruch Tabanpour and Qiayuan Liao and Mustafa Haiderbhai and Samuel Holt and Jing Yuan Luo and Arthur Allshire and Erik Frey and Koushil Sreenath and Lueder A. Kahrs and Carmelo Sferrazza and Yuval Tassa and Pieter Abbeel},
      year={2025},
      eprint={2502.08844},
      archivePrefix={arXiv},
      primaryClass={cs.RO},
      url={https://arxiv.org/abs/2502.08844}, 
}

@misc{brockman2016openaigym,
      title={OpenAI Gym}, 
      author={Greg Brockman and Vicki Cheung and Ludwig Pettersson and Jonas Schneider and John Schulman and Jie Tang and Wojciech Zaremba},
      year={2016},
      eprint={1606.01540},
      archivePrefix={arXiv},
      primaryClass={cs.LG},
      url={https://arxiv.org/abs/1606.01540}, 
}

@techreport{silver2025era,
  title = {Welcome to the Era of Experience},
  author = {David Silver and Richard S. Sutton},
  year = {2025},
  institution = {DeepMind},
  url = {https://storage.googleapis.com/deepmind-media/Era-of-Experience%20/The%20Era%20of%20Experience%20Paper.pdf},
  note = {Preprint of a chapter to appear in the book Designing an Intelligence, MIT Press}
}

@misc{park2023generativeagentsinteractivesimulacra,
      title={Generative Agents: Interactive Simulacra of Human Behavior}, 
      author={Joon Sung Park and Joseph C. O'Brien and Carrie J. Cai and Meredith Ringel Morris and Percy Liang and Michael S. Bernstein},
      year={2023},
      eprint={2304.03442},
      archivePrefix={arXiv},
      primaryClass={cs.HC},
      url={https://arxiv.org/abs/2304.03442}, 
}

@misc{yoon2025omuletorchestratingmultipletools,
      title={OMuleT: Orchestrating Multiple Tools for Practicable Conversational Recommendation}, 
      author={Se-eun Yoon and Xiaokai Wei and Yexi Jiang and Rachit Pareek and Frank Ong and Kevin Gao and Julian McAuley and Michelle Gong},
      year={2025},
      eprint={2411.19352},
      archivePrefix={arXiv},
      primaryClass={cs.AI},
      url={https://arxiv.org/abs/2411.19352}, 
}

@misc{zhou2024webarenarealisticwebenvironment,
      title={WebArena: A Realistic Web Environment for Building Autonomous Agents}, 
      author={Shuyan Zhou and Frank F. Xu and Hao Zhu and Xuhui Zhou and Robert Lo and Abishek Sridhar and Xianyi Cheng and Tianyue Ou and Yonatan Bisk and Daniel Fried and Uri Alon and Graham Neubig},
      year={2024},
      eprint={2307.13854},
      archivePrefix={arXiv},
      primaryClass={cs.AI},
      url={https://arxiv.org/abs/2307.13854}, 
}

@inproceedings{Wang_2025, series={WWW ’25},
   title={Value Function Decomposition in Markov Recommendation Process},
   url={http://dx.doi.org/10.1145/3696410.3714807},
   DOI={10.1145/3696410.3714807},
   booktitle={Proceedings of the ACM on Web Conference 2025},
   publisher={ACM},
   author={Wang, Xiaobei and Liu, Shuchang and Cai, Qingpeng and Li, Xiang and Hu, Lantao and Li, Han and Xie, Guangming},
   year={2025},
   month=apr, pages={379–390},
   collection={WWW ’25} }

@misc{huang2025crmarenaproholisticassessmentllm,
      title={CRMArena-Pro: Holistic Assessment of LLM Agents Across Diverse Business Scenarios and Interactions}, 
      author={Kung-Hsiang Huang and Akshara Prabhakar and Onkar Thorat and Divyansh Agarwal and Prafulla Kumar Choubey and Yixin Mao and Silvio Savarese and Caiming Xiong and Chien-Sheng Wu},
      year={2025},
      eprint={2505.18878},
      archivePrefix={arXiv},
      primaryClass={cs.CL},
      url={https://arxiv.org/abs/2505.18878}, 
}

@misc{deffayet2024sardinesimulatorautomatedrecommendation,
      title={SARDINE: A Simulator for Automated Recommendation in Dynamic and Interactive Environments}, 
      author={Romain Deffayet and Thibaut Thonet and Dongyoon Hwang and Vassilissa Lehoux and Jean-Michel Renders and Maarten de Rijke},
      year={2024},
      eprint={2311.16586},
      archivePrefix={arXiv},
      primaryClass={cs.IR},
      url={https://arxiv.org/abs/2311.16586}, 
}

@misc{xu2025theagentcompanybenchmarkingllmagents,
      title={TheAgentCompany: Benchmarking LLM Agents on Consequential Real World Tasks}, 
      author={Frank F. Xu and Yufan Song and Boxuan Li and Yuxuan Tang and Kritanjali Jain and Mengxue Bao and Zora Z. Wang and Xuhui Zhou and Zhitong Guo and Murong Cao and Mingyang Yang and Hao Yang Lu and Amaad Martin and Zhe Su and Leander Maben and Raj Mehta and Wayne Chi and Lawrence Jang and Yiqing Xie and Shuyan Zhou and Graham Neubig},
      year={2025},
      eprint={2412.14161},
      archivePrefix={arXiv},
      primaryClass={cs.CL},
      url={https://arxiv.org/abs/2412.14161}, 
}

@misc{piao2025agentsocietylargescalesimulationllmdriven,
      title={AgentSociety: Large-Scale Simulation of LLM-Driven Generative Agents Advances Understanding of Human Behaviors and Society}, 
      author={Jinghua Piao and Yuwei Yan and Jun Zhang and Nian Li and Junbo Yan and Xiaochong Lan and Zhihong Lu and Zhiheng Zheng and Jing Yi Wang and Di Zhou and Chen Gao and Fengli Xu and Fang Zhang and Ke Rong and Jun Su and Yong Li},
      year={2025},
      eprint={2502.08691},
      archivePrefix={arXiv},
      primaryClass={cs.SI},
      url={https://arxiv.org/abs/2502.08691}, 
}

@misc{xia2025hierarchicaltreesearchbaseduser,
      title={Hierarchical Tree Search-based User Lifelong Behavior Modeling on Large Language Model}, 
      author={Yu Xia and Rui Zhong and Hao Gu and Wei Yang and Chi Lu and Peng Jiang and Kun Gai},
      year={2025},
      eprint={2505.19505},
      archivePrefix={arXiv},
      primaryClass={cs.IR},
      url={https://arxiv.org/abs/2505.19505}, 
}

@misc{deng2025onerecunifyingretrieverank,
      title={OneRec: Unifying Retrieve and Rank with Generative Recommender and Iterative Preference Alignment}, 
      author={Jiaxin Deng and Shiyao Wang and Kuo Cai and Lejian Ren and Qigen Hu and Weifeng Ding and Qiang Luo and Guorui Zhou},
      year={2025},
      eprint={2502.18965},
      archivePrefix={arXiv},
      primaryClass={cs.IR},
      url={https://arxiv.org/abs/2502.18965}, 
}

@misc{zhai2024actionsspeaklouderwords,
      title={Actions Speak Louder than Words: Trillion-Parameter Sequential Transducers for Generative Recommendations}, 
      author={Jiaqi Zhai and Lucy Liao and Xing Liu and Yueming Wang and Rui Li and Xuan Cao and Leon Gao and Zhaojie Gong and Fangda Gu and Michael He and Yinghai Lu and Yu Shi},
      year={2024},
      eprint={2402.17152},
      archivePrefix={arXiv},
      primaryClass={cs.LG},
      url={https://arxiv.org/abs/2402.17152}, 
}

@misc{hou2025actionpiececontextuallytokenizingaction,
      title={ActionPiece: Contextually Tokenizing Action Sequences for Generative Recommendation}, 
      author={Yupeng Hou and Jianmo Ni and Zhankui He and Noveen Sachdeva and Wang-Cheng Kang and Ed H. Chi and Julian McAuley and Derek Zhiyuan Cheng},
      year={2025},
      eprint={2502.13581},
      archivePrefix={arXiv},
      primaryClass={cs.IR},
      url={https://arxiv.org/abs/2502.13581}, 
}

@misc{wang2024llmsuserexplorationlargescale,
      title={LLMs for User Interest Exploration in Large-scale Recommendation Systems}, 
      author={Jianling Wang and Haokai Lu and Yifan Liu and He Ma and Yueqi Wang and Yang Gu and Shuzhou Zhang and Ningren Han and Shuchao Bi and Lexi Baugher and Ed Chi and Minmin Chen},
      year={2024},
      eprint={2405.16363},
      archivePrefix={arXiv},
      primaryClass={cs.IR},
      url={https://arxiv.org/abs/2405.16363}, 
}

@misc{zhang2025notellm2multimodallargerepresentation,
      title={NoteLLM-2: Multimodal Large Representation Models for Recommendation}, 
      author={Chao Zhang and Haoxin Zhang and Shiwei Wu and Di Wu and Tong Xu and Xiangyu Zhao and Yan Gao and Yao Hu and Enhong Chen},
      year={2025},
      eprint={2405.16789},
      archivePrefix={arXiv},
      primaryClass={cs.IR},
      url={https://arxiv.org/abs/2405.16789}, 
}

@misc{zhu2025longtermclockfinegrainedtime,
      title={Long-Term Interest Clock: Fine-Grained Time Perception in Streaming Recommendation System}, 
      author={Yongchun Zhu and Guanyu Jiang and Jingwu Chen and Feng Zhang and Xiao Yang and Zuotao Liu},
      year={2025},
      eprint={2501.15817},
      archivePrefix={arXiv},
      primaryClass={cs.IR},
      url={https://arxiv.org/abs/2501.15817}, 
}

@misc{huang2025agenticrecommendersystemsera,
      title={Towards Agentic Recommender Systems in the Era of Multimodal Large Language Models}, 
      author={Chengkai Huang and Junda Wu and Yu Xia and Zixu Yu and Ruhan Wang and Tong Yu and Ruiyi Zhang and Ryan A. Rossi and Branislav Kveton and Dongruo Zhou and Julian McAuley and Lina Yao},
      year={2025},
      eprint={2503.16734},
      archivePrefix={arXiv},
      primaryClass={cs.AI},
      url={https://arxiv.org/abs/2503.16734}, 
}

@misc{wang2025gflowgrfinetuninggenerativerecommendation,
      title={GFlowGR: Fine-tuning Generative Recommendation Frameworks with Generative Flow Networks}, 
      author={Yejing Wang and Shengyu Zhou and Jinyu Lu and Qidong Liu and Xinhang Li and Wenlin Zhang and Feng Li and Pengjie Wang and Jian Xu and Bo Zheng and Xiangyu Zhao},
      year={2025},
      eprint={2506.16114},
      archivePrefix={arXiv},
      primaryClass={cs.IR},
      url={https://arxiv.org/abs/2506.16114}, 
}

@misc{chopra2024limitsagencyagentbasedmodels,
      title={On the limits of agency in agent-based models}, 
      author={Ayush Chopra and Shashank Kumar and Nurullah Giray-Kuru and Ramesh Raskar and Arnau Quera-Bofarull},
      year={2024},
      eprint={2409.10568},
      archivePrefix={arXiv},
      primaryClass={cs.MA},
      url={https://arxiv.org/abs/2409.10568}, 
}

@article{Binz2025,
  author = {Binz, Marcel and Akata, Elif and Bethge, Matthias and Brändle, Franziska and Callaway, Fred and Coda-Forno, Julian and Dayan, Peter and Demircan, Can and Eckstein, Maria K. and Éltető, Noémi and Griffiths, Thomas L. and Haridi, Susanne and Jagadish, Akshay K. and Ji-An, Li and Kipnis, Alexander and Kumar, Sreejan and Ludwig, Tobias and Mathony, Marvin and Mattar, Marcelo and Modirshanechi, Alireza and Nath, Surabhi S. and Peterson, Joshua C. and Rmus, Milena and Russek, Evan M. and Saanum, Tankred and Schubert, Johannes A. and Schulze Buschoff, Luca M. and Singhi, Nishad and Sui, Xin and Thalmann, Mirko and Theis, Fabian J. and Truong, Vuong and Udandarao, Vishaal and Voudouris, Konstantinos and Wilson, Robert and Witte, Kristin and Wu, Shuchen and Wulff, Dirk U. and Xiong, Huadong and Schulz, Eric},
  title = {A foundation model to predict and capture human cognition},
  journal = {Nature},
  year = {2025},
  month = {Jul},
  day = {2},
  volume = {},
  number = {},
  pages = {},
  doi = {10.1038/s41586-025-09215-4},
  url = {https://doi.org/10.1038/s41586-025-09215-4},
  issn = {1476-4687}
}

@misc{chen2025pinfmfoundationmodeluser,
      title={PinFM: Foundation Model for User Activity Sequences at a Billion-scale Visual Discovery Platform}, 
      author={Xiangyi Chen and Kousik Rajesh and Matthew Lawhon and Zelun Wang and Hanyu Li and Haomiao Li and Saurabh Vishwas Joshi and Pong Eksombatchai and Jaewon Yang and Yi-Ping Hsu and Jiajing Xu and Charles Rosenberg},
      year={2025},
      eprint={2507.12704},
      archivePrefix={arXiv},
      primaryClass={cs.LG},
      url={https://arxiv.org/abs/2507.12704}, 
}

@misc{zhou2026openonerectechnicalreport,
      title={OpenOneRec Technical Report}, 
      author={Guorui Zhou and Honghui Bao and Jiaming Huang and Jiaxin Deng and Jinghao Zhang and Junda She and Kuo Cai and Lejian Ren and Lu Ren and Qiang Luo and Qianqian Wang and Qigen Hu and Rongzhou Zhang and Ruiming Tang and Shiyao Wang and Wuchao Li and Xiangyu Wu and Xinchen Luo and Xingmei Wang and Yifei Hu and Yunfan Wu and Zhanyu Liu and Zhiyang Zhang and Zixing Zhang and Bo Chen and Bin Wen and Chaoyi Ma and Chengru Song and Chenglong Chu and Defu Lian and Fan Yang and Feng Jiang and Hongtao Cheng and Huanjie Wang and Kun Gai and Pengfei Zheng and Qiang Wang and Rui Huang and Siyang Mao and Tingting Gao and Wei Yuan and Yan Wang and Yang Zhou and Yi Su and Zexuan Cheng and Zhixin Ling and Ziming Li},
      year={2026},
      eprint={2512.24762},
      archivePrefix={arXiv},
      primaryClass={cs.IR},
      url={https://arxiv.org/abs/2512.24762}, 
}

@misc{xing2026reg4recreasoningenhancedgenerativemodel,
      title={REG4Rec: Reasoning-Enhanced Generative Model for Large-Scale Recommendation Systems}, 
      author={Haibo Xing and Hao Deng and Yucheng Mao and Lingyu Mu and Jinxin Hu and Yi Xu and Hao Zhang and Jiahao Wang and Shizhun Wang and Yu Zhang and Xiaoyi Zeng and Jing Zhang},
      year={2026},
      eprint={2508.15308},
      archivePrefix={arXiv},
      primaryClass={cs.IR},
      url={https://arxiv.org/abs/2508.15308}, 
}

@misc{wen2025usbreceffectiveframeworkimproving,
      title={USB-Rec: An Effective Framework for Improving Conversational Recommendation Capability of Large Language Model}, 
      author={Jianyu Wen and Jingyun Wang and Cilin Yan and Jiayin Cai and Xiaolong Jiang and Ying Zhang},
      year={2025},
      eprint={2509.20381},
      archivePrefix={arXiv},
      primaryClass={cs.CL},
      url={https://arxiv.org/abs/2509.20381}, 
}

@misc{bougie2025simusersimulatinguserbehavior,
      title={SimUSER: Simulating User Behavior with Large Language Models for Recommender System Evaluation}, 
      author={Nicolas Bougie and Narimasa Watanabe},
      year={2025},
      eprint={2504.12722},
      archivePrefix={arXiv},
      primaryClass={cs.IR},
      url={https://arxiv.org/abs/2504.12722}, 
}

@inproceedings{parkchoongwon2025,
author = {Park, Choongwon},
title = {LLM as User Simulator: Towards Training News Recommender without Real User Interactions},
year = {2025},
isbn = {9798400715921},
publisher = {Association for Computing Machinery},
address = {New York, NY, USA},
url = {https://doi.org/10.1145/3726302.3730224},
doi = {10.1145/3726302.3730224},
booktitle = {Proceedings of the 48th International ACM SIGIR Conference on Research and Development in Information Retrieval},
pages = {3080–3084},
numpages = {5},
location = {Padua, Italy},
series = {SIGIR '25}
}

@misc{zhang2026exploringrecommenderevaluationmultimodal,
      title={Exploring Recommender System Evaluation: A Multi-Modal User Agent Framework for A/B Testing}, 
      author={Wenlin Zhang and Xiangyang Li and Qiyuan Ge and Kuicai Dong and Pengyue Jia and Xiaopeng Li and Zijian Zhang and Maolin Wang and Yichao Wang and Huifeng Guo and Ruiming Tang and Xiangyu Zhao},
      year={2026},
      eprint={2601.04554},
      archivePrefix={arXiv},
      primaryClass={cs.IR},
      url={https://arxiv.org/abs/2601.04554}, 
}

@misc{liu2026diagnosticguideddynamicprofileoptimization,
      title={Diagnostic-Guided Dynamic Profile Optimization for LLM-based User Simulators in Sequential Recommendation}, 
      author={Hongyang Liu and Zhu Sun and Tianjun Wei and Yan Wang and Jiajie Zhu and Xinghua Qu},
      year={2026},
      eprint={2508.12645},
      archivePrefix={arXiv},
      primaryClass={cs.IR},
      url={https://arxiv.org/abs/2508.12645}, 
}

@misc{chen2025vragentr1boostingvideorecommendation,
      title={VRAgent-R1: Boosting Video Recommendation with MLLM-based Agents via Reinforcement Learning}, 
      author={Siran Chen and Boyu Chen and Chenyun Yu and Yuxiao Luo and Ouyang Yi and Lei Cheng and Chengxiang Zhuo and Zang Li and Yali Wang},
      year={2025},
      eprint={2507.02626},
      archivePrefix={arXiv},
      primaryClass={cs.MM},
      url={https://arxiv.org/abs/2507.02626}, 
}

@misc{he2025plumadaptingpretrainedlanguage,
      title={PLUM: Adapting Pre-trained Language Models for Industrial-scale Generative Recommendations}, 
      author={Ruining He and Lukasz Heldt and Lichan Hong and Raghunandan Keshavan and Shifan Mao and Nikhil Mehta and Zhengyang Su and Alicia Tsai and Yueqi Wang and Shao-Chuan Wang and Xinyang Yi and Lexi Baugher and Baykal Cakici and Ed Chi and Cristos Goodrow and Ningren Han and He Ma and Romer Rosales and Abby Van Soest and Devansh Tandon and Su-Lin Wu and Weilong Yang and Yilin Zheng},
      year={2025},
      eprint={2510.07784},
      archivePrefix={arXiv},
      primaryClass={cs.IR},
      url={https://arxiv.org/abs/2510.07784}, 
}



\clearpage
\includepdf[pages=-,fitpaper=false,pagecommand={\thispagestyle{empty}}]{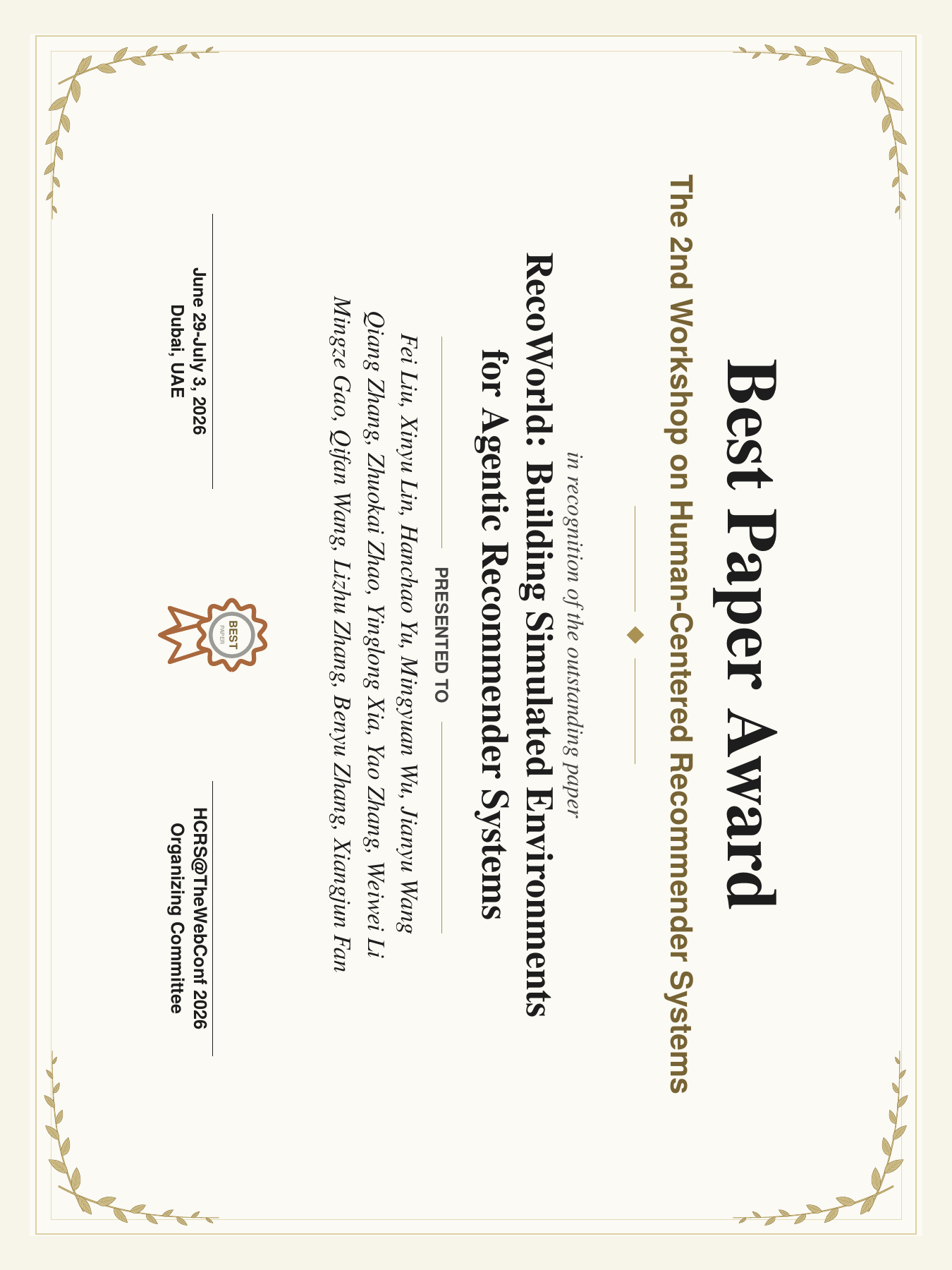}

\end{document}